# Design and Fabrication of Waveguide Optics for Imaging Applications

P. Babu and S. Ganganagunta

Department of Physics, Koneru Lakshmaiah University, Guntur, Andhra Pradesh, India

**Abstract**

Due to the non-ionizing property, researchers have chosen to investigate terahertz radiation (THz) Imaging instrumentation for Bio-Sensing applications. The present work is to design and fabricate a near field lens that can focus guided terahertz radiation to a microscopic region for the detection of cancer-affected cells in Biological tissue. Operational characteristics such as field of view, optical loss factor, and hydrophobicity must be included to achieve an effective design of the lens.

**Introduction**

The recent interest in terahertz technologies has generated a wide range of applications in Bio-Imaging [1-4]. Due to the non-ionizing property, unlike X-rays, terahertz radiation will not damage tissues and DNA like X-rays. Some frequencies of terahertz can penetrate several millimeters of tissue and back reflect the signal to enable detection of the differences in water content and tissue density. According to recent investigation, it is known that cancer effected cells will absorb more water than healthy ones. Therefore, one can find the exact location of effected cells in the tissue through active THz imaging using an endoscope.

An Endoscope is a medical device consisting of a long, thin, flexible tube that comprises of a light source and detector, such as video camera, used for Imaging. Endoscopy is a minimally invasive diagnostic medical procedure used to examine the interior surfaces of an organ. To construct an endoscope for terahertz imaging applications, researchers used flexible waveguides to transmit THz radiation. According to the current research, low loss hollow waveguides for the propagation of terahertz radiation were fabricated by coating silver inside Polycarbonate tubing with diameters on the order of millimeters [5]. The TE$_{11}$ mode was successfully coupled into the waveguides. A loss of 3.5 dB/m was measured for silver coated polycarbonate tube of diameter 4.6 mm [6]. 80% transmission was achieved using silver and polystyrene coated polymer waveguides [7]. These flexible waveguides were determined to have the potential to be used for transporting THz radiation over meter length distances with low transmission losses [8].

These available low loss, hollow, flexible waveguides can be used to transmit the THz radiation. The present problem is to design and fabricate a lens to focus this transmitted THz radiation on to the tissue. Initially a lens will be designed to focus THz radiation.

**Methods of Investigation**

Spherical lenses are not appropriate because of the large beam diameters associated with the Terahertz radiation that results in spherical aberrations (Fig.1). To get rid of spherical aberration we can use parabolic mirrors, as parabolic surface is the only solution for a mirror to convert a plane wavefront into a spherical wavefront.

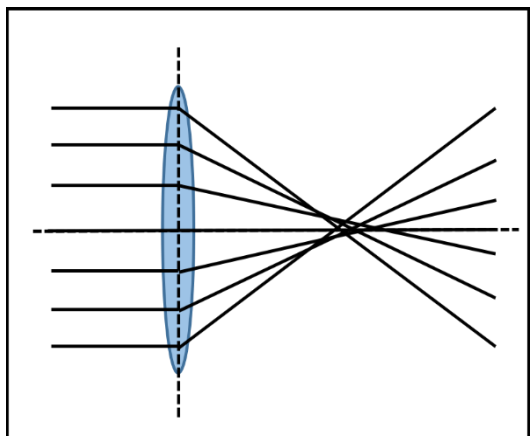

**Figure 1:** Spherical aberration

Though the traditional approach of using a parabolic mirror is diffraction limited, it is susceptible to chromatic aberration (coma) once misaligned as shown in Fig.2. The alignment is always difficult and off axis rays are subject to coma as the direction of the optic axis changes upon reflection off the mirror.

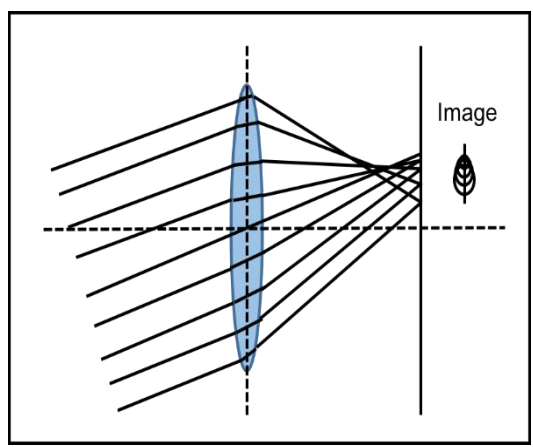

**Figure 2:** Chromatic aberration

Furthermore, the numerical aperture (NA) is limited. For high NAs, the incident beam overlaps with the focal spot. Hence, using a lens with no tilt ensures proper alignment. However, all lenses with their two surfaces will generate different near field patterns and hence the spatial resolution of the system will depend on the lens design.

In order to simplify the calculations for lens design, geometric optics is used to define both surfaces of the lens. It's based on Fermat's principle where the entire wavefront from a collimated incident beam converges into a focal spot and each ray travels the same optical distance. Nevertheless, this does not provide information about focal spot size.

If the focal plane is less than 100λ (λ-Incident wavelength) from the lens, it is essential to use near-field theory to work out intensity distribution in the focal plane, so Kirchhoff's scalar diffraction theory is used to determine the focal spot size.

*The Planar-Hyperbolic (p-h) Lens: -*

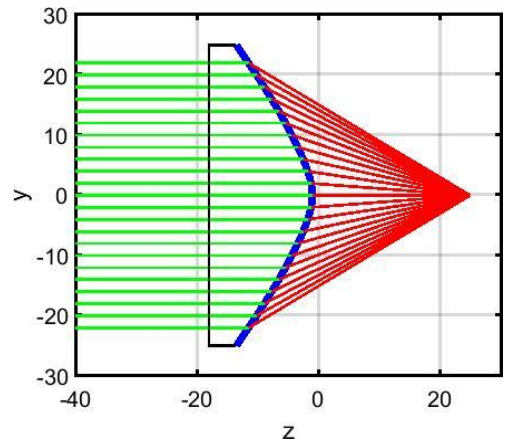

**Figure 3**: Planar-hyperbolic lens

In creating an aberration free lens, as we have the freedom to choose the shape i.e. radius of curvature of each surface of the lens; set the first surface to be flat recognizing all refraction occurs at the second surface. Once the focal length of the lens has been fixed, i.e. radius of the curvature of the second surface, the only design of freedom is the thickness of the lens. The resultant second surface is hyperbola, shown in Fig.3.

The Focal length of a lens in air can be calculated from the *lens maker's equation*:

$$\frac{1}{f} = (n-1)\left[\frac{1}{R_1} - \frac{1}{R_2} + \frac{(n-1)d}{nR_1R_2}\right],$$

Where f is the focal length of the lens, n is the refractive index of the lens material, and R1 and R2 are the radius of curvature of first and second surfaces of the lens.

*Limitation with p-h lens*: The angle between the incident beam and the asymptote to the hyperbola is per definition the critical angle, where the total Internal reflection start.

According to *Snell's Law*,

$$\frac{\sin\theta_1}{\sin\theta_2} = \frac{v_1}{v_2} = \frac{n_2}{n_1}$$

Once the refractive indices $n_1$ and $n_2$ are fixed, then angle of incidence is proportional to the angle of refraction. For lenses with high numerical aperture, the beam will be subjected to large reflection losses since the reflected part of the beam comes back directly from the first surface as it is a flat surface.

*The Elliptical-Aspheric (e-a) Lens:-*

In order to overcome the problem of large reflection losses, we can design the lens with its first surface to be curved as well. By choosing the first surface to be elliptical the second surface must be aspheric to meet the requirement. Here elliptical surface will generate a spherical wavefront with in the lens material and the second surface images this spherical wavefront to another spherical wavefront behind the lens [9], shown in Fig.4.

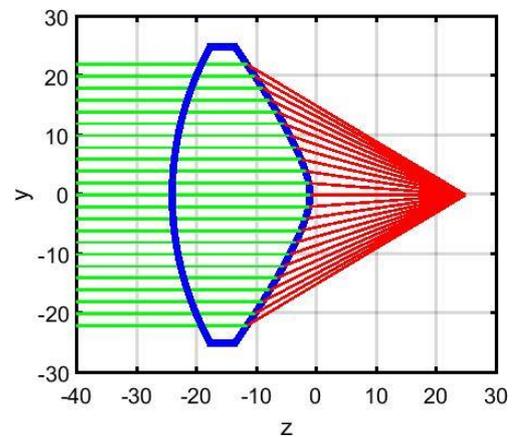

**Figure 4 :** Elliptical-aspheric lens

*Limitation with e-a lens*: Though we can minimize the reflection loss from the second surface due to the total internal reflection by making first surface to be curved as well, it's very difficult to optimize the reflection losses on both surfaces.

*The Symmetric-Pass Lens (s-p lens):-*

To overcome the total internal reflection of the planar-hyperbolic lens, one can minimize the effect by making the first surface to be curved as well and the elliptical-aspheric lens shown in Fig.5 (reflection loss from the both surfaces). We have to choose a lens, which reduces the spherical aberration considerably, such that the incident beam experiences the same angle of deviation on both surfaces while passing through the lens as a result the overall reflection loss would be minimum.

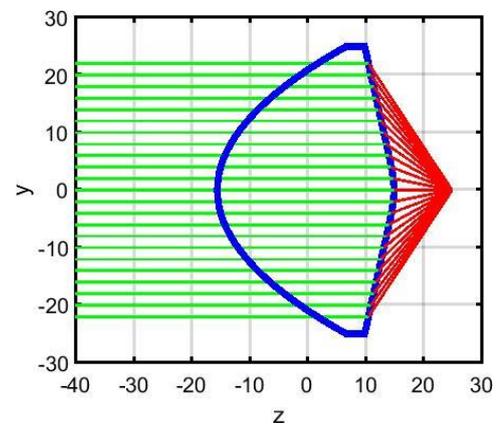

**Figure 5:** Symmetric-pass lens

Due to the long wavelength of terahertz radiation, the spatial resolution (which is directly related to the focal spot size) is quite limited. Based on the resolution, symmetric pass lens was found to be the ideal lens. These lenses are used to get a focal length of 25mm with diameter 50mm for wavelengths of the order 0.1mm. However, symmetric Pass lens is not appropriate for cases where the focal length of the lens needs to be just 3 to 4 times the wave length of the THz radiation. As the wavelength and focal length of the lens are of the same order (~mm), one should use near field optics [11] instead of geometric-optics.

**Ideal Lens Design:**

Ultimately hyper hemi spherical (HHS) lens [10, 12] was chosen to be the ideal candidate for terahertz imaging. Once we design a HHS lens for THz focusing then by making slight modifications to the lens structure one can achieve a lens that can be used for active THz imaging [13-15]. Next step is to minimize optical loss. One will encounter significant power loss while detecting the back signal from the tissue, and it is due to the reflection from flat surface. This can be minimized by coating the surface with the anti-reflecting material with suitable refractive index.

The expression for the Reflection coefficient or Reflectance can be calculated by using Fresnel equations and is valid when the light is at near normal incidence to the surface. According to Lord Raleigh, depositing a thin layer on the surface of lens with a material of refractive index $n_1$ between the index of air $n_0$ and that of lens $n_S$, we can get the optimum value for R.

The relation can be derived as follows,

$$n_1 = \sqrt{n_0 n_S}$$

Fabricating the lens with either Germanium or Quartz and then coating the lens surface with anti-reflective dielectrics like Quartz and Teflon can minimize reflection loss.

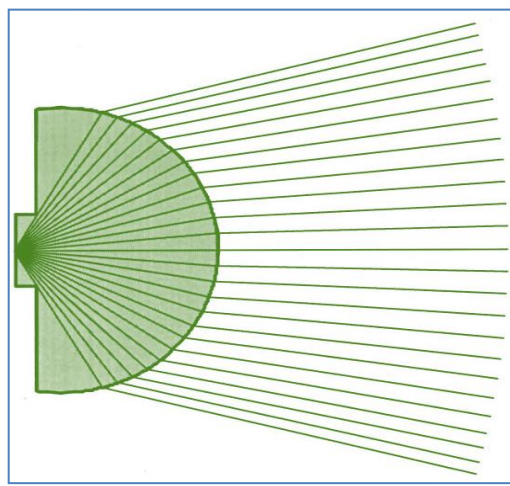

**Figure 6:** Hyper hemi spherical lens for best terahertz performance

**Germanium** with $n_s$=3.91

**Quartz** with $n_s$≈2

**Teflon** with $n_s$=1.4 for THz regime.

$$R = \left(\frac{n_0 - n_S}{n_0 + n_S}\right)^2$$

After reducing the reflection loss, we have to make the surface to be hydrophobic by growing nanostructures using femto second laser ablation so that fluids would not impede the performance of the endoscope, shown in Fig.7.

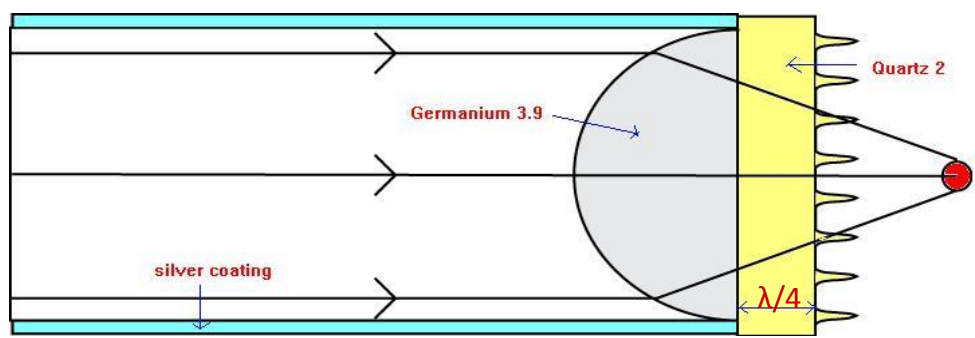

**Figure 7:** Hydrophobic hyper hemi spherical lens attached to metal coated terahertz waveguides for maximum terahertz transmission

**Conclusion:**

Design and fabrication of effective, aberration free lenses with diameter 50mm can focus THz radiation to a distance 25mm by working at wavelengths nearly 0.1mm and have resolution up to the hundreds of micron range.

**References**

1. Doradla, P., Alavi, K., Joseph, C. S., & Giles, R. H. (2014, March). Terahertz polarization imaging for colon cancer detection. In SPIE OPTO (pp. 89850K-89850K). International Society for Optics and Photonics.

2. Ito, T., Matsuura, Y., Miyagi, M., Minamide, H., & Ito, H. (2007). Flexible terahertz fiber optics with low bend-induced losses. JOSA B, 24(5), 1230-1235.

3. Doradla, P., Alavi, K., Joseph, C., & Giles, R. (2013). Detection of colon cancer by continuous-wave terahertz polarization imaging technique. Journal of Biomedical Optics, 18(9), 090504-090504.

4. Doradla, P., Alavi, K., Joseph, C. S., & Giles, R. H. (2013, March). Continuous wave terahertz reflection imaging of human colorectal tissue. In SPIE OPTO (pp. 86240O-86240O). International Society for Optics and Photonics.

5. Kumar, A., Doradla, P., Narkhede, M., Li, L., Samuelson, L. A., Giles, R. H., & Kumar, J. (2014). A simple method for fabricating silver nanotubes. RSC Advances, 4(69), 36671-36674.

6. Doradla, P., Joseph, C. S., Kumar, J., & Giles, R. H. (2012, February). Propagation loss optimization in metal/dielectric coated hollow flexible terahertz waveguides. In SPIE OPTO (pp. 82610P-82610P). International Society for Optics and Photonics.

7. Doradla, P., Joseph, C. S., Kumar, J., & Giles, R. H. (2012). Characterization of bending loss in hollow flexible terahertz waveguides. Optics express, 20(17), 19176-19184.

8. Doradla, P., & Giles, R. H. (2014, March). Dual-frequency characterization of bending loss in hollow flexible terahertz waveguides. In SPIE OPTO (pp. 898518-898518). International Society for Optics and Photonics.

9. Lo, Y. H., & Leonhardt, R. (2008). Aspheric lenses for terahertz imaging. Optics express, 16(20), 15991-15998.

10. Doradla, P., Alavi, K., Joseph, C. S., & Giles, R. H. (2015, April). Flexible waveguide enabled single-channel terahertz endoscopic system. In SPIE OPTO (pp. 93620D-93620D). International Society for Optics and Photonics.

11. Hunsche, S., Koch, M., Brener, I., & Nuss, M. C. (1998). THz near-field imaging. Optics communications, 150(1), 22-26.

12. Doradla, P. (2014). Development of single channel terahetz endoscopic system for cancer detection (Doctoral dissertation, University of Massachusetts Lowell).

13. Doradla, P., Alavi, K., Joseph, C., & Giles, R. (2014). Single-channel prototype terahertz endoscopic system. Journal of biomedical optics, 19(8), 080501-080501.

14. Van Rudd, J., & Mittleman, D. M. (2002). Influence of substrate-lens design in terahertz time-domain spectroscopy. JOSA B, 19(2), 319-329.

15. Doradla, P., Alavi, K., Joseph, C. S., & Giles, R. H. (2016, May). Development of terahertz endoscopic system for cancer detection. In SPIE OPTO (pp. 97470F-97470F). International Society for Optics and Photonics.